\documentclass[journal]{IEEEtran}
\newtheorem{definition}{Definition}
\usepackage{url}
\usepackage{cite}
\usepackage{latexsym}
\usepackage{amsmath}
\usepackage{amstext,amsfonts,epsf,epsfig,pifont}
\usepackage{graphics,graphicx,bm,amsmath,multicol,rotating}

\newlength{\intwidth}

\def\Xint#1{\mathchoice
{\XXint\displaystyle\textstyle{#1}}%
{\XXint\textstyle\scriptstyle{#1}}%
{\XXint\scriptstyle\scriptscriptstyle{#1}}%
{\XXint\scriptscriptstyle\scriptscriptstyle{#1}}%
\!\int}
\def\XXint#1#2#3{{\setbox0=\hbox{$#1{#2#3}{\int}$}
\vcenter{\hbox{$#2#3$}}\kern-.5\wd0}}

\def\dashint{\Xint-}

\newcommand{\bw}{\ensuremath{\mathbf{w}}}
\newcommand{\bx}{\ensuremath{\mathbf{x}}}

\newcommand{\bX}{\ensuremath{\mathbf{X}}}

\newcommand{\calA}{\ensuremath{\mathcal{A}}}

\newcommand{\calH}{\ensuremath{\mathcal{H}}}

\newcommand{\ri}{\ensuremath{i}}

\newcommand{\Imag}{\ensuremath{\Im}}

\begin{document}

\noindent \emph{The following statements are placed here in accordance with the copyright policy of the Institute of Electrical and Electronics Engineers, Inc., available online at}
\url{http://www.ieee.org/publications_standards/publications/rights/rights_policies.html}\\

\noindent
Lilly, J. M., \&  Olhede, S. C. (2011).  Analysis of modulated \\\indent multivariate oscillations. In press at \emph{IEEE Transactions on}\\\indent \emph{Signal Processing}.\\


\noindent \copyright 2011 IEEE. Personal use of this material is permitted. Permission from IEEE must be obtained for all other uses, in any current or future media, including reprinting/republishing this material for advertising or promotional purposes, creating new collective works, for resale or redistribution to servers or lists, or reuse of any copyrighted component of this work in other works.\\

\newpage

\title{Analysis of Modulated Multivariate Oscillations}
\author{Jonathan~M.~Lilly,~\IEEEmembership{Member,~IEEE,}
and Sofia~C.~Olhede,~\IEEEmembership{Member,~IEEE}
\thanks{Manuscript submitted \today.  The work of J. M. Lilly was supported  by awards \#0751697 and \#1031002 from the Physical Oceanography program of the United States National Science Foundation.  The work of S. C. Olhede was supported by award \#EP/I005250/1 from the Engineering and Physical Sciences Research Council of the United Kingdom. }
\thanks{J.~M.~Lilly is with NorthWest Research Associates, PO Box 3027, Bellevue, WA, USA (e-mail: lilly@nwra.com).}
\thanks{S.~C.~Olhede is with the Department of Statistical Science, University College London, Gower Street,
London WC1E 6BT, UK (e-mail: s.olhede@ucl.ac.uk).}}

\markboth{IEEE Transactions on  Signal Processing, Submitted \today}{Lilly \& Olhede: Modulated Multivariate Oscillations}

\maketitle
\begin{abstract}

The concept of a common modulated oscillation spanning multiple time series is formalized, a method for the recovery of such a signal from potentially noisy observations is proposed, and the time-varying bias properties of the recovery method are derived.   The method, an extension of wavelet ridge analysis to the multivariate case, identifies the common oscillation by seeking, at each point in time, a frequency for which a bandpassed version of the signal obtains a local maximum in power.  The lowest-order bias is shown to involve a quantity, termed the instantaneous curvature, which measures the strength of local quadratic modulation of the signal after demodulation by the common oscillation frequency.  The bias can be made to be small if the analysis filter, or wavelet, can be chosen such that the signal's instantaneous curvature changes little over the filter time scale. An application is presented to the detection of vortex motions in a set of freely-drifting oceanographic instruments tracking the ocean currents.
\end{abstract}

\IEEEpeerreviewmaketitle

\section{Introduction}

\IEEEPARstart{I}n the physical sciences the description of common variability in a set of multiple time series is an important data analysis task.  Frequently a key signal present in the data is that of a modulated oscillation, extending across time series channels, with a different amplitude and perhaps a different phase shift in each time series.   Such oscillations may be the signature of waves or wavelike phenomenon.   The most important multivariate cases are the bivariate and trivariate cases, which occur frequently in oceanography and seismology, for example.  One wishes to extract the common oscillatory structure from the observations, a task that is complicated by the possible presence of noise and also by time variability of the signal of interest.

The analysis of univariate modulated oscillations is more highly evolved than is the multivariate case.  In both cases one must begin with a model for the signal structure.  For univariate signals, an attractive representation for an amplitude/frequency modulated signal is now well known, and involves the construction of a complex-valued quantity called the \emph{analytic signal}  \cite{gabor46-piee,vakman72-reep,vakman77-spu,vakman96-itsp,picinbono97-itsp,cohen99-sp}.    Real-world signals are nearly always contaminated by noise or other sources of variability, hence some means of filtering or localizing the time series is required in order to isolate the modulated oscillation.   The analytic signal corresponding to the modulated oscillation of interest can be estimated with a popular and powerful method known as {\em wavelet ridge analysis} \cite{delprat92-itit,mallat,lilly10-itit}. The essence of this method, which is more general than its name might suggest, is a {\em local optimization} applied to a set of frequency-localized versions of the observed time series.

This paper develops a powerful and flexible method, termed {\em multivariate wavelet ridge analysis},  for the extraction of modulated oscillations from multivariate time series. Estimates of the time-varying forms of leading-order bias terms are derived, which are essential in informing the choice of analysis filter or wavelet.  This a non-trivial extension of a related work by the authors for the univariate case \cite{lilly10-itit}.  The key innovation here is a model for signal structure in which a set of signals are expanded in terms of deviations from oscillatory behavior at a single common but time-varying frequency.   The basic idea of  multivariate wavelet ridge analysis, but without a theoretical understanding of the bias, was presented in the preliminary work \cite{lilly09-asilomar}.

The motivation for such a method is the analysis of ocean currents in the now very large set of data from freely-drifting, or ``Lagrangian'', instruments, see e.g. \cite{lapcod} and references therein.   The signatures of a particular type of oceanic structure---long-lived or ``coherent'' vortices \cite{mcwilliams85-rvg}---occur frequently in such data and are aptly described as modulated oscillations in two dimensions.  The development of automated and objective schemes for the analysis of such features has been attempted by several authors \cite{lilly06-npg,flament01-jfm,lankhorst06-jaot}, and thus this work will be of practical value.   An application to a dataset of this type, from the observational experiment  of \cite{richardson89-jpo,armi89-jpo}, is presented here as an illustration.   

The structure of the paper is as follows.  Some essential background is presented in Section~\ref{backgroundsec}, together with a data example.  In Section~\ref{modoscsec} we introduce a representation for a modulated multivariate oscillation, and quantify the degree of variability of such a signal via a local expansion.   A generalization of wavelet ridge analysis appropriate to a multivariate signal is presented in Section~\ref{ridgesection}, and the leading-order bias term is identified.  A key contribution is the identification and interpretation of the quantity controlling the bias, a higher-order relative of the joint instantaneous bandwidth of \cite{lilly10-itsp} which we term the \emph{joint instantaneous curvature}.

All data, numerical algorithms, and functions for analysis and figure generation are distributed to the community as a freely available Matlab package, as described in Appendix~\ref{appendixjlab}.\footnote{This package, called Jlab, is available at \url{http://www.jmlilly.net}.}

\section{Background}\label{backgroundsec}

This section presents the background necessary for the development of an analysis method for treating modulated multivariate oscillations.  A real-world data example of oceanographic data provides a practical motivation.

\subsection{Statement of the Problem}

A set of $N$ real-valued observed time series, assumed square-integrable herein, are arranged as an $N$-vector
\begin{equation}
\bx_o(t)\equiv \left[x_{o;1}(t)\,\,x_{o;2}(t)\,\,\ldots \,\,x_{o;N} (t)\right]^T
\end{equation}
where ``$T$'' denotes the matrix transpose.   At least some of the $N$ channels of $\bx_o(t)$ are expected to contain oscillatory variability, and these oscillations are in turn expected to be related to one another or to share some joint structure.  We therefore model $\bx_o(t)$ as containing two separate components
\begin{equation}\label{signal}
\bx_o(t)=\bx(t)+\bx_r(t)
\end{equation}
where $\bx(t)$ is a {\em modulated multivariate oscillation}, defined subsequently, and $\bx_r(t)$ is a residual which we assume may be accurately represented as a stochastic process.   Thus $\bx(t)$ is the ``signal'' and $\bx_r(t)$ is the ``noise''.  Our goals here are (i) to {\em estimate} the multivariate oscillatory signal $\bx(t)$ given the observed vector $\bx_o(t)$; (ii) to {\em characterize} its time-varying behavior; and to estimate the errors in this process from (iii) {\em bias} associated with $\bx(t)$ itself.   The first step is a model specification for the modulated multivariate oscillation.

\begin{figure*}[]
\begin{center}
\includegraphics[height=7in,angle=-90]{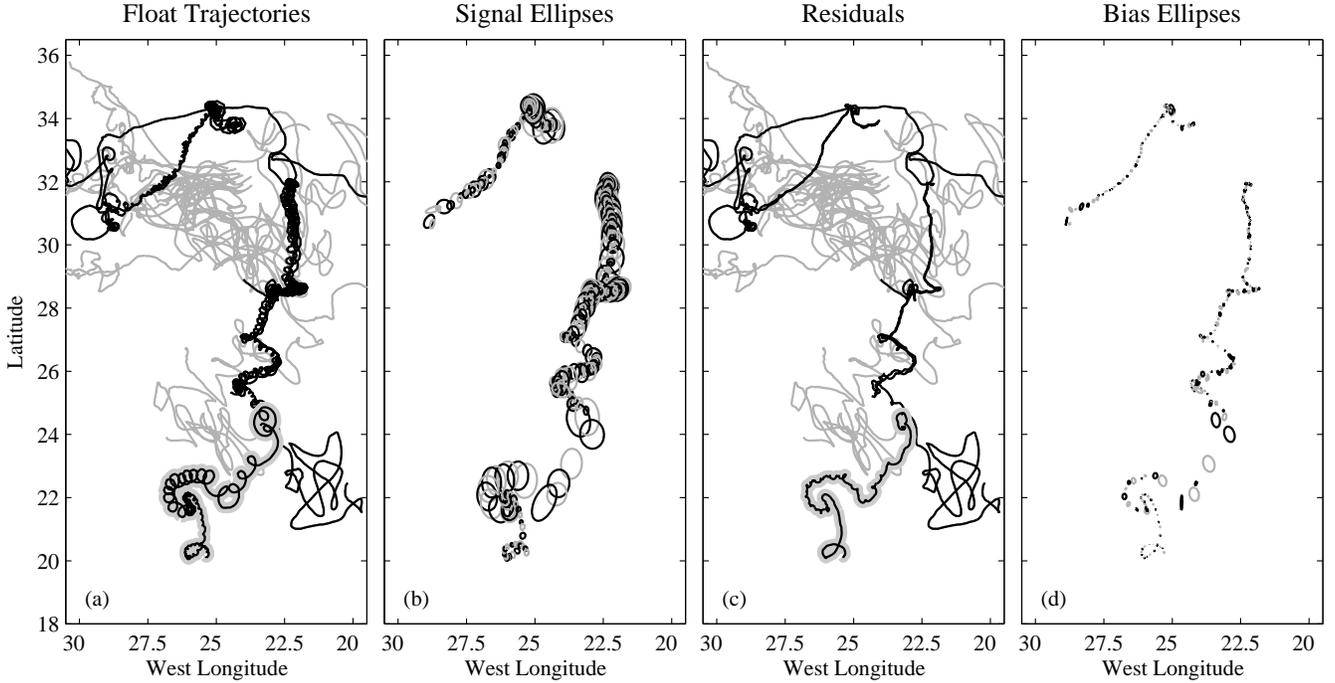}
\end{center}
        \caption{Application of the extraction algorithm for modulated multivariate oscillations to freely-drifting oceanographic instruments from the northeast subtropical Atlantic  \cite{richardson89-jpo,armi89-jpo}.  The observed data $\bx_o(t)$ in (a) is decomposed into a set of  estimated modulated oscillations $\widehat\bx(t)$, shown in (b) as a set of time-varying ellipses, plus a residual $\widehat\bx_r(t)$ in (c).  In (b), ellipses are shown at twice actual size for presentational clarity.  The time interval between the ellipse snapshots varies in time and is equal to the estimated instantaneous period, as described later in the text, with alternating grey and black ellipses.  Panel (d) is the same as (c), but the ellipses represent the instantaneous estimated bias of the signal estimate.  In (a) and (c), twenty-two different records are shown, with black lines used for those records for which a modulated oscillation is found, and grey lines for the remainder. The heavy gray curve in (a) and (c) outlines a particular record that will be used as an example later.       }\label{bandwidth_loopers}
\end{figure*}

\subsection{A Bivariate Example}
An example of data matching the model (\ref{signal}) for the bivariate case of $N=2$ is shown in Fig.~\ref{bandwidth_loopers}, along with our eventual decomposition into an estimated oscillatory portion $\widehat\bx(t)$ plus an estimated residual $\widehat\bx_r(t)$. This data, described in more detail in Section~\ref{application}, is from a set of freely-drifting instruments called ``floats'' that track the ocean currents, recording their horizontal position at regular intervals. Freely-drifting instruments such as these represent one of the primary ways oceanographers study the structure and variability of ocean currents, see e.g. \cite{lapcod} and references therein.

The observed time series $\bx_o(t)$ in Fig.~\ref{bandwidth_loopers}a clearly show the presence of modulated oscillations superposed on a background of apparently random fluctuations, matching the model~(\ref{signal}). The  oscillations in these records represent the presence of long-lived, intense oceanic vortex structures, in this case of about 10--20~km radius \cite{armi89-jpo,richardson89-jpo}.    Vortices such as those seen here are ubiquitous features of the ocean currents [e.g.,~\citen{mcwilliams85-rvg}], and a large number of papers have been devoted to the study of their dynamics and impact on the large-scale flow.   In  Fig.~\ref{bandwidth_loopers}b, the estimated bivariate modulated oscillations $\widehat\bx(t)$ have been visually represented as time-varying ellipses; we refer the reader to \cite{lilly10-itsp} for details on the {\em modulated elliptical signal} representation of a bivariate analytic signal.  Such modulated elliptical signals are a special case of a more general class of signals we will consider; it is worth pointing out that ellipses form the building blocks for models of many different kinds of data, from ocean currents \cite{gonella72-dsr} to seismic signals \cite{park87b-jgr}  to electroencephalographic (EEG) data~\cite{allefeld09-chaos}.

 \subsection{Fundamentals}

The starting point for our analysis is the {\em analytic signal method} \cite{gabor46-piee,cohen99-sp,vakman72-reep,vakman77-spu,vakman96-itsp,picinbono97-itsp} for assigning meaningful time-varying amplitudes and frequencies to each of the channels of $\bx(t)$.  The {\em analytic part} of the signal vector is defined as
\begin{equation}
\bx_+(t)\equiv 2\calA \,\bx(t)\equiv \bx(t)+\ri \calH\bx(t)
\label{analyticsignal}
\end{equation}
where ``$\calH$'' denotes the Hilbert transform operator
\begin{equation}
\calH\bx(t)
\equiv \frac{1}{\pi}\,\dashint_{-\infty}^\infty\frac{ \bx(\tau)}{ t-\tau}\,d  \tau
\label{hilbert}
\end{equation}
with ``$\dashint$'' being the Cauchy principal value integral.   The Fourier transforms of $\bx(t)$ and $\bx_+(t)$ are denoted $\bX(\omega)$ and $\bX_+(\omega)$, respectively.  It follows from the form of the analytic operator $\calA$ in the frequency domain that
\begin{equation}
\bX_+(\omega)= 2U(\omega) \bX(\omega)\label{analyticfreqdomain}
\end{equation}
where $U(\omega)$ is the unit step function.  Thus the application of the operator $2\calA$ to $\bx(t)$ doubles  the amplitudes of the Fourier coefficients  of $\bx(t)$ at positive frequencies, while causing the coefficients at negative frequencies to vanish.

A set of $N$ unique amplitudes $a_n(t)$ and phases $\phi_n(t)$ is then implicitly defined by
\begin{equation}
\bx_+(t)=\begin{bmatrix} 2\calA x_1(t)\\  2\calA x_2(t)\\ \vdots\\ 2\calA x_N(t)
   \end{bmatrix}\equiv \begin{bmatrix}a_1(t) e^{\ri\phi_1(t)}\\ a_2(t) e^{\ri\phi_2(t)}\\\vdots\\ a_N(t) e^{\ri\phi_N(t)}
   \end{bmatrix}\label{multivariateanalytic}
\end{equation}
with the amplitudes being non-negative, $a_n(t)\ge 0$.  The $n$th amplitude $a_n(t)$ and phase $\phi_n(t)$ constructed in this manner are called the \emph{canonical} amplitude and phase associated with the $n$th signal channel $x_n(t)$.  Taking the real part, $\bx(t)=\Re\left\{\bx_+(t)\right\}$, each signal channel is now described as a modulated oscillation with time-varying amplitude $a_n(t)$ and phase $\phi_n(t)$.   The derivative of the $n$th phase, $\omega_n(t)\equiv\phi_n'(t)$, is called the $n$th {\em instantaneous frequency} \cite{gabor46-piee,boashash92a-ieee,picinbono97-itsp}, which gives the local frequency of oscillation of the $n$th signal.

The analytic signal method provides the foundation for describing each channel of $\bx(t)$ as an oscillation with time-varying properties.  While the assignment of an amplitude and a phase to a given real-valued signal cannot be unique, the compelling properties of the amplitude and phase derived from the analytic signal are now well known \cite{gabor46-piee,cohen99-sp,vakman72-reep,vakman77-spu,vakman96-itsp}; see \cite{picinbono97-itsp} for a useful review.   Since a wide variety of physical processes can be aptly described as a set of modulated oscillations, the representation of the multivariate signal $\bx(t)$ as in (\ref{multivariateanalytic}) is a strongly motivated and powerful model.

\section{Modulated Multivariate Oscillations}\label{modoscsec}

In this section the notion of a modulated multivariate oscillation is formalized.  The key is a local expansion of the signal about a demodulated version of itself.  This expansion quantifies the signal's departure, at each moment, from the best possible fit to a set of sinusoidal oscillations all sharing a single frequency, i.e. from a {\em pure oscillation}.   First-order and second-order deviations are introduced which quantify instantaneous linear and quadratic modulation, and which play a central role in an aggregate description of the signal's variability.

\subsection{A Local Signal Expansion}\label{section:expansion}

The joint evolution of the multivariate signal $\bx(t)$ in the vicinity of time $t$ may be locally represented in terms of a series of deviations from a set of constant-amplitude oscillations all evolving with some common instantaneous frequency $\omega(t)$.   With $\tau$ representing a time offset or ``local time'', the analytic signal may be written in the vicinity of a reference time $t$ as
\begin{multline}\label{simplex}
\bx_+(t+\tau)= e^{i\omega(t)\tau}
\left\{\bx_+(t)+\tau\,\widetilde\bx_1(t;\omega(t)) \right.\\\left.+\frac{1}{2}\tau^2\widetilde\bx_2(t;\omega(t)))  +\bm{\epsilon}_3(t,\tau;\omega(t)))\right\}
\end{multline}
a representation we refer to as the \emph{local modulation expansion}.  The local modulation expansion describes the evolution of a multivariate signal as being due to the phase progression at a {\em single time-varying frequency} $\omega(t)$, together with a series of deviations from this behavior.  This model of joint structure is a key contribution, since it represents the multivariate signal as a single object, rather than as a set of unrelated oscillations.

The $p$\,th vector-valued coefficient of the expansion, termed the $p$\,th-order {\em deviation vector}, is given by
\begin{equation}
\widetilde\bx_p(t;\omega(t))   \equiv \frac{\partial^p}{\partial\tau^p}\left.\left[ e^{-i\omega(t)\tau}\bx_+(t+\tau)\right]\right|_{\tau=0}\label{curvdef}
\end{equation}
while the remainder term takes the form
\begin{equation}
\bm{\epsilon}_\bx(t,\tau;\omega(t))\label{brdef}
\equiv\frac{1}{6}\tau^3\frac{\partial^3}{\partial\tau^3}\left.\left[ e^{-i\omega(t)\tau}\bx_+(t+\tau)\right]\right|_{\tau=u}
\end{equation}
for some (unknown) point $u$ contained in the interval $\left[0,\tau\right]$, as follows from the Lagrange form of the remainder in the Taylor series \cite[p~880]{abramowitz}.
To derive (\ref{simplex}), write
\begin{equation}
\bx_+(t+\tau) =  e^{i\omega(t)\tau}\left[ e^{-i\omega(t)\tau}\bx_+(t+\tau) \right]
\end{equation}
and then Taylor-expand the term in square brackets with respect to  the point $\tau=0$.

The local modulation expansion (\ref{simplex}) states that the lowest-order joint behavior of the signal $\bx_+(t+\tau)$, considered as a function of local time $\tau$ in the vicinity of a fixed reference time~$t$, is for all signal channels to undergo a phase progression at a single frequency $\omega(t)$.  The next-order behavior is a linear tendency in $\tau$, controlled by the first deviation vector $\widetilde\bx_1(t;\omega(t))$, which is also subject to the phase progression at frequency $\omega(t)$.  The still next-order behavior is a quadratic tendency in $\tau$, controlled by the second deviation vector $\widetilde\bx_2(t;\omega(t))$.  Since $\widetilde\bx_1(t;\omega(t))$ and $\widetilde\bx_2(t;\omega(t))$ are complex-valued in general, they impact both the amplitudes and phases of the oscillations in the various signal channels.  In the vicinity of times $t$ for which the signal $\bx(t)$ is usefully described as a modulated oscillation at frequency $\omega(t)$, the remainder term  $\bm{\epsilon}_\bx(t,\tau;\omega(t))$ is expected to be negligible provided $\tau$ is not too large compared with the oscillation period $2\pi/\omega(t)$.

\subsection{The Best Fit Frequency}\label{section:extremization}

Our aim is to describe the common or joint oscillatory structure of the signal $\bx(t)$.   To this end, note that the quantity
\begin{multline}
\upsilon_\bx^2(t;\omega(t))\equiv\frac{\left\|\widetilde\bx_1(t;\omega(t))\right\|^2}{\|\bx_+(t)\|^ 2}\\=\frac{\left\|\bx_+'(t)-i\omega(t)\bx_+(t)\right\|^2}{\|\bx_+(t)\|^ 2}
\label{multivariatefrequencyerror}
\end{multline}
is the normalized instantaneous error involved in locally approximating the rate of change of the analytic version of $\bx(t)$ as undergoing a uniform phase progression with some local frequency $\omega(t)$.   For example, with $x_+(t)=a_o e^{i\omega_o t}$, $x_+'(t)-i\omega_o x_+(t)$ vanishes.  One way to determine the best choice of frequency in the local modulation expansion  (\ref{simplex}) is therefore to find that $\omega(t)$ which minimizes the error $\upsilon_\bx^2(t;\omega(t))$.  Differentiating (\ref{multivariatefrequencyerror}) with respect to $\omega(t)$ at each time~$t$ gives
\begin{equation}
\frac{1}{2}\frac{\partial}{\partial[\omega(t)]}\upsilon_\bx^2(t;\omega(t))= -\frac{\Imag\left\{\bx_+^H(t)\bx_+'(t)\right\}}{\|\bx_+(t)\|^ 2}+\omega(t)
\label{multivariatefrequencyerrorfull}
\end{equation}
and we see, upon setting this quantity equal to zero, that an extremum in the fractional error $\upsilon_\bx^2(t;\omega(t))$ occurs for
\begin{equation}
\omega_\bx(t)\equiv\frac{\Imag\left\{\bx_+^H(t)\bx_+'(t)\right\}}{\|\bx_+(t)\|^ 2}=
\frac{\sum_{n=0}^N a_n^2(t) \omega_n(t)} {\sum_{n=0}^N a_n^2(t)}
\label{multivariatefrequency}
\end{equation}
which is the {\em power-weighted average} of the~$N$ component frequencies $\omega_n(t)$. The second derivative of (\ref{multivariatefrequencyerror}) is positive at this value of $\omega(t)$, so this extremum is in fact a minimum.

Thus $\omega_\bx(t)$ defined in (\ref{multivariatefrequency})  minimizes the leading-order deviation vector in the local modulation expansion (\ref{simplex}), and is in a sense the ``best fit'' local frequency.  The expression (\ref{multivariatefrequency}) has in fact been encountered before, in \cite{lilly10-itsp}.  Therein it was shown that the power-weighted time average of $\omega_\bx(t)$ satisfies an important global constraint---it recovers the first moment of the channel-averaged Fourier spectrum of $\bx_+(t)$---and thus $\omega_\bx(t)$ generalizes the concept of ``instantaneous frequency''  \cite{gabor46-piee,boashash92a-ieee,picinbono97-itsp} to the multivariate case.  That this  {\em joint instantaneous frequency} $\omega_\bx(t)$ also has a compelling {\em local} interpretation as the solution to a  minimization problem is another reason why it is a natural measure of the common time-varying frequency content of~$\bx(t)$.  The interpretation of the instantaneous frequency as the solution to a local minimization problem holds for the standard univariate instantaneous frequency, since $\omega_x(t)=\phi_x'(t)$ for $x_+(t)=a_x(t)e^{i\phi_x(t)}$ is the $N=1$ special case of the joint instantaneous frequency.

Henceforth we choose $\omega(t)$ in  the local modulation expansion (\ref{simplex}) to take the value $\omega(t)=\omega_\bx(t)$, that is, we write
\begin{multline}\label{simplexomegax}
\bx_+(t+\tau)= e^{i\omega_\bx(t)\tau}\times\\
\left\{\bx_+(t)+\tau\,\widetilde\bx_1(t) +\frac{1}{2}\tau^2\widetilde\bx_2(t)  +\bm{\epsilon}_\bx(t,\tau)\right\}
\end{multline}
where the deviation vectors and residual are defined as
\begin{align}
\widetilde\bx_p(t)&\equiv\widetilde\bx_p(t;\omega_\bx(t)) \\
\bm{\epsilon}_\bx(t,\tau)&\equiv\bm{\epsilon}_\bx(t,\tau;\omega_\bx(t)).
\end{align}
We refer to the $\widetilde\bx_p(t)$ as the {\em intrinsic deviation vectors}, since the demodulation can be seen as a sort of coordinate transformation, with the natural or intrinsic choice of coordinate system being the one in which the phase progression follows the  joint instantaneous frequency.

The first two intrinsic deviation vectors are central in understanding the time-dependent joint structure of $\bx(t)$ as a modulated oscillation.  These are explicitly given by \begin{align}
\widetilde\bx_1(t)&=\bx_+'(t)-i\omega_\bx(t)\bx_+(t)\label{upexpress}\\
\widetilde\bx_2(t) & = \bx_+''(t)-\ri2\omega_\bx(t)\bx_+'(t)-\omega_\bx^2(t)\bx_+(t)\label{curvexpress}
\end{align}
the right-hand sides of which are oscillator equations that describe the first-order and second-order departure, respectively, of the evolution of $\bx_+(t)$ from a local oscillation at the frequency $\omega_\bx(t)$.  The magnitudes these vectors, compared with the signal strength, are quantified by
\begin{equation}
\upsilon_\bx(t) \equiv \frac{\left\|\widetilde\bx_1(t)\right\|}{\|\bx_+(t)\|},\quad\quad
\xi_\bx(t) \equiv \frac{\left\|\widetilde\bx_2(t)\right\|}{\|\bx_+(t)\|}
\label{rhodef}
\end{equation}
which will occur frequently in what follows.   The first of these was also encountered in \cite{lilly10-itsp}, in which it was shown that $\upsilon_\bx^2(t)$ gives the time-varying contribution to the second central moment of the channel-averaged spectrum of $\bx_+(t)$ that is not explained by variations of the joint instantaneous frequency $\omega_\bx(t)$ about its time-mean value.  Thus $\upsilon_\bx(t)$ is called the {\em joint instantaneous bandwidth} and is the natural multivariate generalization of the univariate instantaneous bandwidth introduced by  \cite{cohen,cohen89-ieee,cohen88-spie}.   The local modulation expansion (\ref{simplexomegax}) shows that the joint instantaneous bandwidth  $\upsilon_\bx(t)$ has a compelling local interpretation as the magnitude of the leading-order deviation of the multivariate signal $\bx_+(t)$ from oscillatory behavior.

\subsection{Physical Interpretation}

It is helpful at this point to say some words about the interpretation of the vectors that have been encountered.   If $\bx(t)$ is taken to represent a position, then $\bx'(t)$ is a velocity and $\bx''(t)$ is an acceleration, and $\bx_+(t)$, $\bx_+'(t)$, and $\bx_+''(t)$, are the {\em analytic parts} of the position, velocity, and acceleration vectors, respectively.\footnote{Note that since differentiation and the analytic operator commute, the analytic part of a derivative is the same as the derivative of the analytic part.}    Then $\widetilde\bx_1(t)$ could be termed the {\em intrinsic analytic velocity}, that is, that part of the analytic velocity which remains if the phase progression at the joint instantaneous frequency is removed, and $\widetilde\bx_2(t)$ could be termed the {\em intrinsic analytic acceleration}.  It turns out that the third derivative of position, $\bx'''(t)$, has an accepted name: it is called the {\em jerk}, see e.g. \cite{schot78-ajp}.  Thus the remainder  $\bm{\epsilon}_\bx(t,\tau)$ occurring in (\ref{simplexomegax})  is the supremum of the {\em intrinsic analytic jerk}.  Constraining $\bm{\epsilon}_\bx(t,\tau)$ to be small therefore amounts to a kind of smoothness condition, namely, that the demodulated analytic signal does not exhibit too much jerkiness in its evolution.  The fact that such smoothness is reasonable to expect for signals that may usefully be considered to be modulated oscillations is an argument in favor of  our truncation of the local modulation expansion  (\ref{simplexomegax})  at the quadratic term.

\subsection{The Deviation Vectors}

In this section we look at the deviation vectors in more detail.  The squared norms of the deviation vectors take simple forms in the univariate case $x_+(t)=a_x(t)e^{i\phi_x(t)}$.  Then (\ref{upexpress}) for $\upsilon_\bx(t)$ becomes $\left|a_x'(t)/a_x(t)\right|$, which is recognized as the modulus of the univariate instantaneous bandwidth \cite{loughlin00-jfi,loughlin00-ssap,loughlin00-ispl}.   Squaring  (\ref{curvexpress}) for $\xi_\bx(t)$ gives in the univariate case
\begin{equation}\label{univariatecurvature}
   \xi_x^2(t)
  =\left[\frac{a_x''(t)}{a_x(t)}\right]^2+\left[\omega_x'(t)\right]^2= \left|\frac{a_x''(t)}{a_x(t)} + i \omega_x'(t)\right|^2
\end{equation}
which involves a squared second derivative of both the amplitude $a_x(t)$ and the phase $\phi_x(t)$, since $\omega_x(t)\equiv\phi'_x(t)$.  The complex-valued quantity $a_x''(t)/a_x(t) + i \omega_x'(t)$, the squared magnitude of which occurs in (\ref{univariatecurvature}), has been previously identified as the coefficient of $\tau^2$ in the local modulation expansion of a univariate signal \cite{lilly10-itit}.   A reasonable name for this univariate quantity is the {\em instantaneous curvature}.  Then $\xi_\bx(t)$ for $N>1$ would be termed the {\em joint instantaneous curvature}, since it quantifies variability which has the same effect that amplitude curvature $a_x''(t)$ and phase curvature $\phi_x''(t)$ have in the univariate case.

In what follows we will need some results concerning the deviation vectors.    The first is that
\begin{equation}\label{parallelfirstderivreal}
\frac{\bx_+^H(t)\widetilde\bx_1(t)}{\|\bx_+(t)\|^ 2}  =  \frac{\|\bx_+(t)\|'}{\|\bx_+(t)\|}
\end{equation}
and thus the quantity on the left-hand-side is real-valued.  This states that the component of the first deviation vector along the direction specified by the signal vector is the fractional rate of change of the total signal amplitude.  In the univariate case, this becomes the bandwidth $a_x'(t)/a_x(t)$.  To see (\ref{parallelfirstderivreal}), project the analytic signal onto its own first derivative to give\begin{equation}\label{parallelfirstderiv}
\frac{\bx_+^H(t)\bx_+'(t)}{\|\bx_+(t)\|^ 2}  =  \frac{\|\bx_+(t)\|'}{\|\bx_+(t)\|}+i\omega_\bx(t)
\end{equation}
using the definition (\ref{multivariatefrequency}) of $\omega_\bx(t)$ together with
\[\|\bx_+(t)\|'=\frac{d}{dt}\sqrt{\bx_+^H(t)\bx_+(t)}=\Re\left\{\bx_+^H(t)\bx_+'(t)\right\}/\|\bx_+(t)\|\]
and then (\ref{parallelfirstderivreal})  follows from  $\widetilde\bx_1(t)=\bx_+'(t)-i\omega_\bx(t)\bx_+(t)$.  

The second result we will need is that $\xi_\bx(t)\ge |\omega_\bx'(t)|$, that is, that the instantaneous curvature is greater than the magnitude of the instantaneous chirp rate.    To derive this, note that the derivative of the left-hand side of (\ref{parallelfirstderiv}) is
\begin{multline}
\frac{d}{dt}\left[ \frac{\bx_+^H(t)\bx_+'(t)}{\|\bx_+(t)\|^ 2} \right]=
\frac{\bx_+^H(t)\bx_+''(t)}{\|\bx_+(t)\|^ 2} + \frac{\|\bx_+'(t)\|^2}{\|\bx_+(t)\|^ 2}\\ -
2 \frac{\|\bx_+(t)\|'}{\|\bx_+(t)\|}
\frac{\bx_+^H(t)\bx_+'(t)}{\|\bx_+(t)\|^ 2} 
\end{multline}
but we may also difference the right-hand side of  (\ref{parallelfirstderiv}) to find
\begin{equation}\label{parallelfirstderiv1}
\frac{d}{dt}\left[ \frac{\bx_+^H(t)\bx_+'(t)}{\|\bx_+(t)\|^ 2} \right]=
\frac{d}{dt} \left[\frac{\|\bx_+(t)\|'}{\|\bx_+(t)\|}\right]+i\omega_\bx'(t).
\end{equation}
The imaginary parts of these two expression combine to give
\begin{equation}
\Im\left\{\frac{\bx_+^H(t)\bx_+''(t)}{\|\bx_+(t)\|^ 2}\right\}=\omega_\bx'(t) +2\omega_\bx(t) \frac{\|\bx_+(t)\|'}{\|\bx_+(t)\|}
\end{equation}
and using (\ref{curvexpress})  to eliminate $ \bx_+''(t)$ together with (\ref{parallelfirstderiv}), we find
\begin{equation}
\Im\left\{\frac{\bx_+^H(t)\widetilde\bx_2(t)}{\|\bx_+(t)\|^ 2}\right\}=\omega_\bx'(t).
\end{equation}
Now introducing the component of $\widetilde\bx_2(t)$ projected onto the signal vector $\bx_+(t)$ as
\begin{equation}
\widetilde\bx_{2,\parallel}(t)\equiv\frac{\bx_+^H(t)\widetilde\bx_2(t)}{\|\bx_+(t)\|^ 2}\bx_+(t)
\end{equation}
we may apply the  Cauchy-Schwarz inequality, leading to
\begin{equation}
\xi_\bx^2(t) \ge  \frac{\left\|\widetilde\bx_{2,\parallel}(t)\right\|^2}{\|\bx_+(t)\|^2} \ge   \frac{\left\|\Im\left\{\widetilde\bx_{2,\parallel}(t)\right\}\right\|^2}{\|\bx_+(t)\|^2}
\end{equation}
and hence $\xi_\bx(t)\ge |\omega_\bx'(t)|$, as stated.

\subsection{Pure Oscillations and Phase Signals}\label{section:modulation}

To understand the distinction between the linear and quadratic terms in the local modulation expansion, we introduce two particularly simple types of multivariate oscillatory signals.  A signal $\bx(t)$ may be said to be a {\em multivariate pure oscillation} if its analytic part is given by
\begin{equation}
\bx_+(t) = e^{i \omega_o t} \bx_o
\end{equation}
for some fixed vector $\bx_o$ and fixed frequency $\omega_o$.  Similarly $\bx(t)$ may be termed a  {\em multivariate phase signal} if
\begin{equation}
\bx_+(t) = e^{i \phi_\bx(t)} \bx_o \label{phasesignal}
\end{equation}
for a fixed vector $\bx_o$ and analytic phase function $e^{i \phi_\bx(t)}$.  The multivariate phase signal (\ref{phasesignal}) is the natural generalization of the univariate phase signal of e.g.~\cite{picinbono97-itsp}. A univariate phase signal may be written as $x_+(t)=|a_o| e^{i\phi_x(t)}$ where the $|a_o|$ is the signal amplitude and $e^{i \phi_x(t)}$ is analytic. In the multivariate case, $\bx_o$ is complex-valued in general as it incorporates information on phase shifts between channels. Thus the constant part $\bx_o$ of the phase signal  can only be interpreted as an amplitude for the univariate case $N=1$, in which case it can be made real-valued and nonnegative by absorbing its phase into $e^{i\phi_x(t)}$.

Phase signals of the form (\ref{phasesignal})  are frequency modulated, with each signal channel having identical time-varying instantaneous frequency $\omega_n(t)=\omega_\bx(t)\equiv\phi_\bx'(t)$, but they are not amplitude modulated since all of the $N$ amplitudes are constant.   Phase signals are the more general class since all pure oscillations are phase signals but not vice-versa.   Note that there are strong constraints on the class of phase functions $\phi_\bx(t)$ such that $ e^{i \phi_\bx(t)}$ be analytic; see e.g. the detailed discussion of univariate phase signals in \cite{picinbono97-itsp}.

The intrinsic deviation vectors take very simple forms for these two types of signals.  For a pure oscillation, $\widetilde\bx_p(t)$ vanishes identically for all $p>0$.  For a phase signal, we~have
\begin{align}
\widetilde\bx_1(t) &=0\label{doesvanish}\\
\widetilde\bx_2(t) &=i\omega_\bx'(t)\bx_+(t)\label{doesntvanish}
\end{align}
so that the first deviation vector vanishes, but the second deviation vector is nonzero whenever the joint instantaneous frequency varies with time.   The former expression (\ref{doesvanish}) follows directly from (\ref{upexpress}).  The latter (\ref{doesntvanish}) may be readily found by rewriting  (\ref{curvexpress}) for the second deviation vector~as
\begin{equation}\label{xtilde1deriv}
\widetilde\bx_2(t)= i\omega_\bx'(t)\bx_+(t)+\left[\widetilde\bx_1'(t)- i\omega_\bx(t)\widetilde\bx_1(t)\right]
\end{equation}
where the first term depends on the {\em joint chirp rate} $\omega_\bx'(t)$, and the other terms vanish when $\widetilde\bx_1(t)$ vanishes.

This illustrates a subtle distinction between the linear and quadratic terms in the local modulation expansion.  Both pure oscillations and phase signals have vanishing {\em linear} deviations from local oscillatory behavior at all times, as measured by the norm of the first deviation vector $\widetilde\bx_1(t)$.  However, phase signals differ from pure oscillations at second order, since the former have non-vanishing {\em quadratic}  deviations from local oscillatory behavior.   Conversely, when $\widetilde\bx_1(t)$ is negligible in the vicinity of time~$t$, we may say that the signal locally evolves as if it were a phase signal having a frequency $\omega_\bx(t)$.  When  $\widetilde\bx_2(t)$ is also negligible, we may say that the signal behaves as a pure oscillation up to second order.  When the leading-order term $e^{i\omega_\bx(t)\tau}\bx_+(t)$ dominates in  (\ref{simplexomegax})  for $\tau$ not too large, we may say that the signal evolves in the vicinity of time $t$ as would be expected for a pure oscillation having frequency $\omega_\bx(t)$.

\subsection{Definition of  a Modulated Multivariate Oscillation}
We are now in a position to formalize what is meant by a modulated oscillation in an arbitrary number of dimensions.   This is accomplished by proposing a single measure of the degree of departure of a multivariate signal from a pure oscillation.  An $N$-channel real-valued zero-mean signal $\bx(t)$ is assumed to have an analytic version $\bx_+(t)$ that is defined and thrice differentiable over some time interval~$T$.    The analytic signal $\bx_+(t)$ is then expanded via the local modulation expansion (\ref{simplexomegax}) using the joint instantaneous frequency $\omega_\bx(t)$.

\begin{definition}{The Modulated Multivariate Oscillation}\\
\label{modmult}
Let  the {\em local stability level} $\delta_T$ be the smallest positive constant  satisfying  for all $t\in T$ the constraints
\begin{equation}
\label{cond1}
\left|\frac{\upsilon_\bx(t)}{\omega_\bx(t) }\right|\le \delta_T ,\quad \quad
\left|\frac{\xi_\bx(t)}{\omega_\bx^2(t)}\right|\le \delta_T^2 
\end{equation}
together with\begin{equation}\label{jerk}
\sup_{t \in T}\frac{1}{\left|\omega_\bx(t)\right|^3}\frac{\|\bm{\epsilon}_\bx(t,
\tau)\|}{\|\bx_+(t)\|}\le \delta^3_T.
\end{equation}
Strongly modulated signals correspond to large values of $\delta_T$, while $\delta_T$ vanishes for a pure oscillation.
The signal $\bx(t)$ is said to be a {\em modulated multivariate oscillation} over time interval $T$ if  the local stability level is less than unity, $\delta_T<1$.
\end{definition}

In this definition, modulated multivariate oscillations occupy a continuum---classified according the local stability level $\delta_T$---with pure oscillations as the limiting or ideal case of vanishing modulation strength.  Signals for which $\delta_T$ exceeds unity present temporal variability that locally exceeds the rate of change of phase at least somewhere on the time interval~$T$.  Such extremely strong modulation would be evidence that the signal is not well modeled in terms of an oscillation at a common time-varying frequency.  An important point is that the class of modulated multivariate oscillations is far larger than that of the so-called ``asymptotic'' signals, see e.g. the discussion in \cite{delprat92-itit}, which   roughly correspond to univariate signals having negligible modulation strength, $\delta_T\ll1$.

In the next section, this ability to quantify the degree of variability becomes essential for determining the time-varying bias involved in the recovery a modulated oscillation from a noisy observation.

\section{Multivariate wavelet ridge analysis}\label{ridgesection}

In this section, a local optimization method---{\em multivariate wavelet ridge analysis}---is created that is able to extract estimates of a modulated multivariate oscillation from potentially noisy observations. Time-varying forms for the leading-order bias effects are also derived.  This extends the work of \cite{delprat92-itit} and \cite{mallat} on the univariate wavelet ridge method, and that of \cite{lilly10-itit} on its bias properties, to the multivariate case.

\subsection{Wavelet Basics}

To isolate a signal of interest from surrounding variability, a time/frequency localized filter is necessary.   A  \emph{wavelet} $\psi(t)$ is a square-integrable complex-valued function satisfying the admissibility
condition \cite{Holschneider}
\begin{equation}
\int_{-\infty}^\infty \frac{\left|\Psi(\omega)\right|^ 2}{\left|\omega\right|}\, d \omega<  \infty
\end{equation}
and the wavelet is said to be {\em analytic} if its Fourier transform $\Psi (\omega)\equiv \int\psi(t)\, e^{-i \omega  t }\,d t $ vanishes for all negative frequencies. The wavelet transform of a real-valued square-integrable vector-valued signal $\bx(t)$ with respect to the wavelet $\psi(t)$ is
\begin{align}
\label{vectorwavetrans}
\bw_{\bx,\psi}(t ,s) & \equiv  \int_{-\infty}^{\infty} \frac{1}{s}
  \psi^*\left(\frac{\tau-t}{s}\right)\,\bx(\tau)\,d \tau\\
    &= \frac{1}{2\pi}\int_{0}^{\infty} \Psi^*(s\omega) \bX(\omega)\, e^{i\omega t}\,d \omega\label{fourierform}
\end{align}
where the latter form follows by the convolution theorem.  With the $1/s$ normalization in (\ref{vectorwavetrans}) rather than the more common $1/\sqrt{s}$, the  wavelet transform can be seen as a set of bandpass operations indexed by the scale~$s$.   Considered as a function of time at each scale $s$, the wavelet transform is seen as a stack of analytic signals.   The frequency-domain wavelet $ \Psi(\omega) $ obtains a maximum modulus at a frequency $\omega_\psi$ called the \emph{peak frequency}.   Without loss of generality, we set $\Psi (\omega_\psi) = 2$.  With these choices, the wavelet transform $w_{x,\psi}(t,s)$ of a sinusoid $x(t)=a_1\cos(\omega_1t+\phi_1) $ obtains a maximum modulus at the scale $s=\omega_\psi/\omega_1$, and the value of this maximum recovers the amplitude of the sinusoid, $|w_{x,\psi}(t ,\omega_\psi/\omega_1)|=|a_1|$.

\subsection{Joint Ridges of a Multivariate Signal}\label{estimationsection}

The detection of a modulated oscillation within the analytic wavelet transform $\bw_{\bx,\psi}(t,s)$ is accomplished as follows.  A {\em ridge point} of $\bw_{\bx,\psi}(t ,s)$ is a time/scale pair $\left(t,s\right)$ satisfying
\begin{equation}
\frac{\partial}{\partial s}\, \left\|\bw_{\bx,\psi}(t ,s)\right\| =  0\label{ampvectorridge},\quad \quad
\frac{\partial^2}{\partial s^ 2}\,  \left\|\bw_{\bx,\psi}(t ,s)\right\| <   0.
\end{equation}
Thus ridge points are locations where the norm of the wavelet transform vector achieves a local maximum with respect to scale.  This definition of a multivariate ridge point\footnote{This type of ridge point is called an ``amplitude ridge point'' by \cite{lilly10-itit}.  Another possibility is a ``phase ridge point'', utilizing a stationary phase condition as proposed by \cite{mallat}.  However, since  \cite{lilly10-itit} find negligible difference between the two types of ridge points in a perturbation analysis, and give practical reasons to prefer the amplitude ridge points, only these will be considered here.} is the natural generalization of the definition for the univariate case\cite{lilly10-itit}, as proposed in the earlier preliminary work~\cite{lilly09-asilomar}.

Adjacent ridge points are then connected to each other to yield a single-valued, continuous function of time called a {\em ridge curve} $\widehat s(t)$ that extends over some time interval $T$.  In practice, two numerical thresholds must be introduced:  a bound on the magnitude of  $\frac{d}{dt}\widehat s(t)$, to avoid jumps across scale, and a minimum ridge duration, to avoid spurious ridges that are short compared with the wavelet length.  Having identified a ridge curve $\widehat s(t)$, the {\em ridge-based estimate} of the analytic signal is then given by
\begin{equation}
\widehat \bx_{+}(t)\equiv \bw_{\bx,\psi}\!\left( t, \widehat s(t)\right),\quad t\in  T\label{amplitudeestimator}
\end{equation}
which is simply the set of values taken by the wavelet transform along the ridge curve.     In order for the wavelet ridge estimate $\widehat \bx_{+}(t)$ to be a good estimate, it is necessary to choose the wavelet properties to match the signal properties.   This is addressed in the next section.

\begin{figure}
\begin{center}
\includegraphics[height=3.5in,angle=-90]{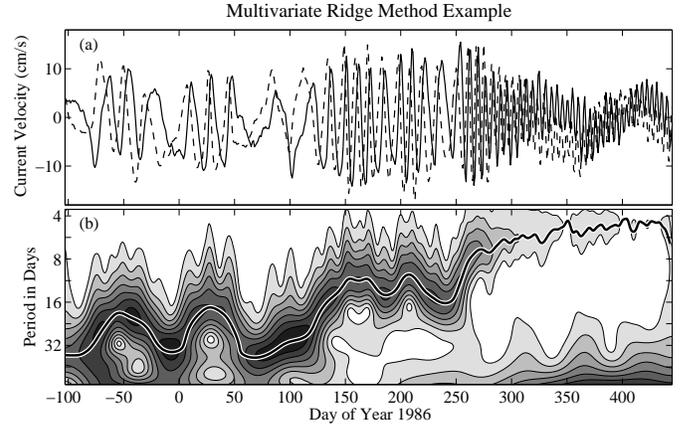}
\end{center}
            \caption{A bivariate position signal, differentiated in time for presentational clarity, is plotted in (a). The solid curve represent eastward velocity and the dashed curve represents northward velocity.  The modulus of the wavelet transform $\bw_{\bx,\psi}(t,s)$ shown in (b), has units of kilometers and is plotted with a logarithmic $y$-axis. The contours range from 0 to 65 km with a spacing of 5~km. The heavy curve is a single unbroken ridge resulting from the application of the multivariate ridge algorithm.  This time series is from the position signal marked by the heavy gray curve in Fig.~\ref{bandwidth_loopers}a and Fig.~\ref{bandwidth_loopers}c.
            }\label{bandwidth_example}
\end{figure}

An example of the multivariate wavelet ridge algorithm is shown in Fig.~\ref{bandwidth_example}.   One of the bivariate time series from Fig.~\ref{bandwidth_loopers}---specifically, that trajectory which is marked by the heavy gray curve in Fig.~\ref{bandwidth_loopers}a and Fig.~\ref{bandwidth_loopers}c---is presented together with its wavelet transform using a choice of wavelet and parameter settings to be discussed later in Section~\ref{application}.  The wavelet transform modulus $\|\bw_{\bx,\psi}(t,s)\|$ shows a clear maximum value as a function of scale, expressed here as the period $2\pi s/\omega_\psi$.  The scale at which this maximum value occurs changes considerably throughout the record, decreasing by an order of magnitude from the beginning to the end as well as presenting some low-frequency variability.   The multivariate ridge curve $\widehat s(t)$ is seen to follow the variability of the maximum of $\|\bw_{\bx,\psi}(t,s)\|$ as a function of time. Evaluating the wavelet transform along this time-dependent curve as in  (\ref{amplitudeestimator}) defines the estimated modulated oscillation $\widehat \bx_{+}(t)$, which is plotted in Fig.~\ref{bandwidth_loopers} as a set of time-varying ellipses together with the estimated residual $\widehat \bx_r(t)\equiv \bx_o(t)-\Re\left\{\widehat \bx_{+}(t)\right\}$.

In the following we will use a measure of the distance, at time $t$, of a scale point $s$ from the instantaneous frequency curve $\omega_\bx(t)$, called the {\em scale deviation}
 \begin{equation}\label{neighborhooddef}
 \Delta\omega_{\bx,\psi}(t,s) \equiv \frac{s\omega_\bx(t)}{\omega_\psi}-1.
\end{equation}
On the time-varying scale curve $s(t)=\omega_\psi/\omega_\bx(t)$ corresponding to the instantaneous frequency curve $\omega_\bx(t)$, the scale deviation vanishes.  One may envision that the constraint $| \Delta\omega_{\bx,\psi}(t,s) | \le |c|$ for some small $|c|>0$ paints out a swath surrounding the instantaneous frequency curve, with the width of this swath increasing as $|c|$ increases.  If $| \Delta\omega_{\bx,\psi}(t,s) | \le \delta_T^2$,  the scale point $s$ is said to lie in the {\em neighborhood} of the instantaneous frequency curve at time $t$. In the next sections we find the conditions under which the ridge equations (\ref{ampvectorridge}) have a solution within the instantaneous frequency neighborhood.

\subsection{Constraints on the Wavelet}

The most important wavelet parameter after its peak frequency $\omega_\psi$ is the {\em dimensionless duration} $P_\psi$, defined by
 \begin{equation}
P_\psi\equiv 
\sqrt{-\omega_\psi ^ 2\frac{\Psi''(\omega_\psi)}{\Psi(\omega_\psi)}}.
\end{equation}
The quantity under the radical is positive for a wavelet with a real-valued Fourier transform $\Psi(\omega)$, since the wavelet then obtains a maximum value at $\omega_\psi$, making $\Psi''(\omega_\psi)$ negative.   It may be shown that $P_\psi/\pi$ corresponds to the number of oscillations at period $2\pi/\omega_\psi$ that fit within the central window of the time-domain wavelet \cite{lilly09-itsp}.  Also $1/P_\psi$ is seen as a dimensionless measure of the wavelet bandwidth, since a Taylor expansion of $\Psi(\omega)$ about the peak frequency gives
\begin{equation}
\Psi(\omega)\approx\Psi(\omega_\psi)\left[1-\frac{1}{2}\left(\frac{\omega/\omega_\psi-1}{1/P_\psi}\right)^2 \right],\quad\omega\approx\omega_\psi\label{pbandwidth}.
\end{equation}
If $P_\psi=1$, the half-power points in this quadratic approximation occur at zero and $2\omega_\psi$, and so the frequency support is extremely broad. One thus expects $P_\psi\ge1$ for wavelet functions that are usefully localized in the frequency domain.

More generally, the basic features of the wavelet may be characterized by its \emph{dimensionless derivatives} \cite{lilly09-itsp}
\begin{equation}\label{dimensionlessdefinition}
\widetilde \Psi_{p}(\omega)\equiv \omega ^ p\frac{\Psi^{(p)}(\omega)}{\Psi(\omega)}
\end{equation}
which are then evaluated at the peak frequency $\omega_\psi$.  Note that $\widetilde \Psi_{1}(\omega_\psi)$ vanishes by definition and that $P_\psi^2=-\widetilde \Psi_{2}(\omega_\psi)$.  The  \emph{wavelet suitability criteria} \cite{lilly10-itit}
\begin{align}
\frac{1}{p!}\left|\widetilde\Psi_p(\omega_\psi)\right|& \le \left\{\begin{array}{lcl}
\delta^{-p/2}_T  &&\quad \frac{p}{2}\in{\mathbb{Z}}\\
\delta^{-(p-1)/2}_T  & &\quad \frac{p+1}{2}\in{\mathbb{Z}}\label{waveletstability}
\end{array} \right.
\end{align}
are a set of conditions that limit the size of the wavelet's dimensionless derivatives at the peak frequency $\omega_\psi$, compared with the inverse of the signal stability level~$\delta_T$.  As the signal becomes more rapidly varying, $\delta_T$ increases, and these bounds on the size of the dimensionless derivatives becomes more stringent.  Note that the second condition in (\ref{waveletstability}) places a stronger constraint on odd derivatives compared to the even derivatives, a reasonable constraint that also proves convenient for the subsequent analysis; see \cite{lilly10-itit} for details on this choice.

It will be seen shortly that the wavelet suitability conditions are the key to ensuring that the bias of the ridge-based signal estimate remains small.  The lowest-order suitability condition, at $p=2$,  implies the wavelet duration is $P_\psi\le\sqrt{2/\delta_T}$.  This means that the more strongly the multivariate signal   $\bx(t)$ varies over an oscillation period---reflected by an increasing value of $\delta_T$---the fewer oscillations the wavelet can contain in time.  But as the wavelet becomes narrower in time it must become broader in frequency, and combining this with the earlier discussion, one expects $1\le P_\psi \le \sqrt{2/\delta_T}$ for wavelets that are usefully localized in both domains.  If $\delta_T$ were to become large compared to unity, no such wavelet could be found.  Our definition of a modulated multivariate oscillation implies $\delta_T<1$, ensuring that the suitability conditions may be satisfied for a wavelet with $P_\psi$ in the range $1\le P_\psi \le \sqrt{2}$.

\subsection{The Wavelet Transform of a Multivariate Oscillation}

The wavelet transform $\bw_{\bx,\psi}(t,s)$ can be expanded into a hierarchy of terms that reveal the interaction between the signal variability and the wavelet shape, allowing the lowest-order bias term to be identified.    A change of variables applied to the wavelet transform (\ref{vectorwavetrans}), and substituting $\bx(t)=\left[\bx_+(t)+\bx_+^*(t)\right]/2$, leads to the form
\begin{equation}\label{simplew}
\bw_{\bx,\psi}(t ,s)  =\frac{1}{2}\int_{-\infty}^{\infty}
  \!\psi^*(\tau)\,\bx_+(t+s\tau)\,d  \tau
\end{equation}
where the contribution of $\bx_+^*(t+\tau)$ to the integrand vanishes on account of the analyticity of the wavelet, as is clear form the Fourier-domain form (\ref{fourierform}).  Inserting the local modulation expansion~(\ref{simplexomegax}) of $\bx_+(t)$, we obtain
\begin{multline}
\bw_{\bx,\psi}(t ,s)=\frac{1}{2}\int_{-\infty}^{\infty}\label{wavelettransformexpansion}
  \!\psi^*\left(\tau\right)\, e^{i\omega_\bx(t)s\tau}\times\\\left\{\bx_+(t)+s\tau\,\widetilde\bx_1(t)+\frac{1}{2}\,(s\tau)^2\,\widetilde\bx_2(t)\right\}\,d \tau
\\+\Delta\bw_{\bx,\psi}(t,s).
\end{multline}
Note that to guarantee the square-integrability of the $\tau^2$ term in the modulation expansion, the long-time decay of the wavelet must be $|\psi(t)|/|\psi(0)|\sim |t|^{-r_\psi}$ for some number $r_\psi\ge 3$.

There are some subtleties surrounding the residual term $\Delta\bw_{\bx,\psi}(t,s)$.  This is implicitly defined as the difference between the left-hand side of (\ref{wavelettransformexpansion}) and the integral on the right-hand side.   It is not the same as the wavelet transform of the Taylor-series remainder $\bm{\epsilon}_\bx(t,\tau)$, because the form (\ref{brdef}) for $\bm{\epsilon}_\bx(t,\tau)$ is only valid over the interval $T$ where we have assumed the signal is differentiable.  Bounding the residual term  $\Delta\bw_{\bx,\psi}(t,s)$ has been examined in the univariate case by  \cite{lilly10-itit}, and since there is no major difference in the multivariate case, we refer the reader there for a detailed discussion.  In general we may expect this term to be very small when the time decay of the wavelet is stronger than $t^{-3}$ for signals that meet our definition of modulated multivariate oscillations.

Assuming the suitability criteria  (\ref{waveletstability}) are satisfied, the wavelet transform in the instantaneous frequency neighborhood, $\Delta\omega_{\bx,\psi}(t,s)=O(\delta_T^2)$, takes the simple form
\begin{multline}\label{readytoexpand3}
\bw_{\bx,\psi}(t,s) = \bx_+(t)- \overset{O(\delta_T)}{\overbrace{\frac{1}{2}\widetilde\Psi_2^*(\omega_\psi)\frac{\widetilde\bx_2(t)}{\omega_\bx^2(t)}}}\\
-\overset{O(\delta_T^2)}{\overbrace{i\Delta\omega_{\bx,\psi}(t,s)\widetilde\Psi_2^*(\omega_\psi)\frac{\widetilde\bx_1(t)}{\omega_\bx(t)}}}
+O(\delta_T^3)+\Delta\bw_{\bx,\psi}(t,s)
\end{multline}
as will be proved shortly.   This powerful result states that at scale points sufficiently close to instantaneous frequency curve, the wavelet transform approximately recovers the analytic signal $\bx_+(t)$. The time-varying forms of the two lowest-order deviations from the analytic signal, up to second order in $\delta_T$, are explicitly resolved.

The derivation of  (\ref{readytoexpand3}) is as follows.  Note that the $p$th frequency-domain derivative of the wavelet is given by
\begin{equation}
\Psi^{(p)}(\omega) = \int_{-\infty}^\infty (-i\tau)^p\psi(\tau) e^{-i\omega\tau} d\tau.
\end{equation}
Evaluating the conjugate of this quantity along the time- and scale-varying frequency $\omega=s\omega_\bx(t)$, we may define a joint function of the wavelet and the signal as
\begin{multline}
\Phi_p(t,s) \equiv \left[s \omega_\bx(t)\right]^p \frac{1}{2}\left[\Psi^{(p)}(s\omega_\bx(t))\right]^*=\\
\left[s \omega_\bx(t)\right]^p\frac{1}{2} \int_{-\infty}^{\infty}\left(i\tau\right)^p  \psi^*(\tau) \, e^ {i s\omega_\bx(t) \tau} \,d \tau\label{fdef}
\end{multline}
which is a function of the time-scale plane.   Inserting (\ref{fdef}) into the wavelet transform expression   (\ref{wavelettransformexpansion}) leads to
\begin{multline}
\bw_{\bx,\psi}(t ,s)=\Phi_0(t,s)\bx_+(t) -i \Phi_1(t,s)\frac{ \widetilde\bx_1(t)}{\omega_\bx(t)}
\\-\frac{1}{2}\Phi_2(t,s) \frac{\widetilde\bx_2(t)}{\omega_\bx^2(t)}+\Delta\bw_{\bx,\psi}(t,s)\label{readytoexpand}
\end{multline}
in which the first two deviation vectors appear explicitly.  This can be simplified by finding approximate expressions for the $\Phi_p(t,s)$ that are valid in the instantaneous frequency neighborhood.

In terms of its  dimensionless derivatives  $\widetilde \Psi_{p}(\omega)$,  the wavelet has a Taylor expansion about the peak frequency $\omega_\psi$ of
\begin{equation}
 \Psi(s\omega)=\Psi(\omega_\psi)\sum_{k= 0}^\infty \frac{1}{k!} \widetilde \Psi_{k}(\omega_\psi) \left(\frac{s\omega}{\omega_\psi}-1\right)^k\label{normalisedexpansionwavelet}.
\end{equation}
 Differentiating both sides, and recalling $\Psi(\omega_\psi)=2$, we find
\begin{equation}
 \frac{1}{2}\Psi^{(p)}(s\omega)= \frac{1}{\omega_\psi^p}\sum_{k= 0}^\infty \frac{1}{k!} \widetilde \Psi_{k+p}(\omega_\psi) \left(\frac{s\omega}{\omega_\psi}-1\right)^k\label{normalisedexpansion}
\end{equation}
for the Taylor series expansion of the $p$\,th derivative of the wavelet, after employing a change in the index of summation.\footnote{Note that (\ref{normalisedexpansion}) corrects the similar expression (114) of \cite{lilly10-itit}.   The latter is only {\em approximately} correct, up to order $\delta_T^2$. Subsequent perturbation expansions in  \cite{lilly10-itit} are not affected because the erroneous contributions occur at unresolved orders in $\delta_T$.}  Thus (\ref{fdef}) becomes, after making use of (\ref{neighborhooddef}),  
\begin{multline}
\Phi_p(t,s)=\left[1+\Delta\omega_{\bx,\psi}(t,s)\right]^p\times\\ \sum_{k= 0}^\infty \frac{1}{k!} \widetilde \Psi_{k+p}^*(\omega_\psi) \left[\Delta\omega_{\bx,\psi}(t,s)\right]^k\label{normalisedexpansion2}
\end{multline}
in terms of the scale deviation.  This may be approximated in the instantaneous frequency neighborhood by making use of the wavelet suitability conditions, and recalling that $\Delta\omega_{\bx,\psi}(t,s)$ is $O(\delta_T^2)$ in the instantaneous frequency neighborhood by definition. The first five $\Phi_p(t,s)$ are found to be
\begin{align}\label{f}
\Phi_0(t,s) &= 1 + O (\delta_T^3) =O(1)\\
\Phi_1(t,s) &= \Delta\omega_{\bx,\psi}(t,s)\widetilde\Psi_2^*(\omega_\psi)+ O (\delta_T^3)= O (\delta_T)   \\
\Phi_2(t,s) &=\widetilde\Psi_2^*(\omega_\psi) +O(\delta_T)= O \left(\delta_T^{-1}\right)\label{f3}\\
\Phi_3(t,s) &= \widetilde\Psi_3^*(\omega_\psi) +O(1) =O \left(\delta_T^{-1}\right)\label{f4}\\
\Phi_4(t,s) &=\widetilde\Psi_4^*(\omega_\psi) +O(\delta_T) =O \left(\delta_T^{-2}\right)\label{f5}
\end{align}in the instantaneous frequency neighborhood.  Using (\ref{f})--(\ref{f3}) and gathering terms by order in (\ref{readytoexpand}), the result (\ref{readytoexpand3}) follows.

\subsection{The Bias of the Wavelet Ridge Method}

One may now show that the ridge equations have a solution within the instantaneous frequency neighborhood. That is, there exists a scale curve
\begin{equation}\label{scalecurveform}
\widehat s(t)=\frac{\omega_\psi}{\omega_\bx(t)}\left[1+O(\delta_T^2) \right]
\end{equation}
which satisfies the ridge equations (\ref{ampvectorridge}).  The proof of this statement, which extends a similar result for univariate case by \cite{lilly10-itit} to the multivariate case, is given in Appendix~\ref{appendixscalederiv}.   Then within the instantaneous frequency neighborhood, (\ref{readytoexpand3}) applies, and we obtain
\begin{multline}
\widehat\bx_\psi(t)\equiv\bw_{\bx,\psi}\left(t, \widehat s(t)\right)
=\bx_+(t)\\ +\frac{1}{2}P_\psi^2\frac{\widetilde\bx_2(t)}{\omega_\bx^2(t)}+O(\delta_T^2)+
\Delta\bw_{\bx,\psi}\left(t,\omega_{\bx,\psi}/\omega_\bx(t)\right)\label{localisedvector}
\end{multline}
as an explicit form for the estimated signal $\widehat\bx_\psi(t)$, resolving the lowest-order time-dependent bias term.

There are several important implications of this result.  The leading error term is due to the \emph{second} deviation vector, and not the first deviation vector.  This is attractive since it means that the lowest-order deviation of the signal from a modulated oscillation---a linear tendency in local time~$\tau$---does not impact the analysis errors at lowest order.  This leading-order error is seen to be associated with the value of the wavelet transform along the instantaneous frequency curve, with the deviation of the ridge from the instantaneous frequency curve only contributing at higher orders in $\delta_T$.   A measure of the total error of the signal estimate is the norm of the difference between the original signal and the estimate, normalized by the signal amplitude, found to be
\begin{equation}\label{errorapprox}
\frac{\|\widehat\bx_\psi(t)-\bx_+(t)\|}{\|\bx_+(t)\|}\approx\frac{1}{2}P_\psi^2  \frac{| \xi_\bx(t)|}{\omega_\bx^2(t)} 
\end{equation}
 which is controlled by the joint instantaneous curvature.  Since $P_\psi$ is the wavelet duration, this states that the error is proportional to the degree of signal curvature over the time support of the wavelet.  To make the leading-error bias term negligible, one must be able to choose the wavelet duration such that this term is sufficiently small.

The joint instantaneous frequency $\omega_\bx(t)$ may also be estimated.  One possibility is to form an estimate by substituting the signal estimate $\widehat\bx_\psi(t)$ for the true signal $\bx_+(t)$ in the definition of the instantaneous frequency (\ref{multivariatefrequency}).  However, following \cite{lilly10-itit}, we instead form the {\em joint transform frequency}
\begin{equation}\label{transformfreq}
\Omega_{\bx,\psi}(t,s)\equiv\frac{\Im\left\{\bw_{\bx,\psi}^H(t,s)\frac{\partial}{\partial t}\bw_{\bx,\psi}(t,s)\right\}}{\left\|\bw_{\bx,\psi}(t,s)\right\|^2}
\end{equation}
and evaluate this quantity along the ridge to obtain the {\em ridge-based instantaneous frequency estimate}
\begin{equation}
\widehat\omega_{\bx,\psi}(t)\equiv \Omega_{\bx,\psi}(t,\widehat s(t)).
\end{equation}
In Appendix~\ref{appendixscalederiv}, we find
\begin{multline}
\widehat\omega_{\bx,\psi}(t)=\omega_\bx(t)\left[
1 -\frac{1}{2}P_\psi^2
 \frac{\Im\left\{\widetilde\bx_1^H(t)\widetilde\bx_2(t)-\bx_+^H(t)\widetilde\bx_3(t)\right\}}{\|\bx_+(t)\|^2\,\omega_\bx^3(t)}\right]\\
= \omega_\bx(t)\left[1 + O(\delta_T^2)\right]
\end{multline}
as an expression for the time-dependent form of this instantaneous frequency estimate.  The wavelet suitability conditions ensure that this estimate is accurate to second order in the local stability level $\delta_T$.

The bias itself may be similarly estimated.  We form the a version of second deviation vector associated with the wavelet transform
\begin{multline}
\widetilde \bw_{2;\bx,\psi}(t,s)\equiv \frac{\partial^2}{\partial t^2}\bw_{\bx,\psi}(t,s)-i 2\Omega_\psi(t,s)\frac{\partial}{\partial t}\bw_{\bx,\psi}(t,s)\\- \Omega_\psi^2(t,s)\bw_{\bx,\psi}(t,s)
\end{multline}
which is created by substituting $\bw_{\bx,\psi}(t,s)$ for $\bx_+(t)$ in  (\ref{curvexpress}), replacing total time derivatives with partial time derivatives.  Then we have the estimates
\begin{equation}
\widehat{\widetilde \bx}_{2;\psi}(t) \equiv \widetilde \bw_{2;\bx,\psi}(t,\widehat s(t)), \quad \quad \widehat\xi_{\bx,\psi}(t)\equiv \frac{\left\|  \widehat{\widetilde \bx}_{2;\psi}(t)\right\|}{\|\widehat{\bx}_{\psi}(t) \|}
\end{equation}
for the second deviation vector and its modulus, the joint instantaneous curvature.   This permits the bias of the estimated signal, and the normalized bias magnitude in (\ref{errorapprox}), to be estimated.

\section{Application}\label{application}

This section illustrates the multivariate wavelet ridge method with an application to real-world bivariate data.  The data is from a set of  instruments tracking the ocean currents, and is representative of a large amount of similar oceanographic data, see e.g. \cite{lapcod}.

\subsection{Data}

The data, shown earlier in Fig.~\ref{bandwidth_loopers}a, consists of position records from twenty-two freely-drifting acoustically-tracked subsurface floats.  These were deployed off the west coast of Africa in the eastern North Atlantic in order to study the local currents in an early experiments of this type \cite{richardson89-jpo,armi89-jpo}.  The instruments are designed to remain neutrally buoyant near a particular depth, 1000 meters in this case, and are tracked acoustically by triangulating sound travel times between the instruments and nearby fixed points.   In the experiment shown here, the sample rate was one day, and only float records with a length exceeding 200 days are presented.  This dataset and many similar ones are available from the World Ocean Circulation Experiment Subsurface Float Data Assembly Center.\footnote{\url{http://wfdac.whoi.edu}}

\subsection{Choice of Wavelet Family}
In implementing the wavelet ridge analysis, the choice of family of analytic wavelets emerges as being important to obtaining desirable properties of the transform, an issue that  has been investigated in detail by \cite{lilly09-itsp} and \cite{lilly10-itit}.   A particularly attractive choice is the generalized Morse wavelet family  \cite{daubechies88-ip,olhede02-itsp,lilly09-itsp}, given by the frequency-domain from
\begin{equation}
\Psi_{\beta,\gamma}(\omega)=U(\omega) a_{\beta,\gamma}\, \omega^\beta e^{-\omega^\gamma}
\end{equation}
where $U(\omega)$ is again the unit step function, $a_{\beta,\gamma}$ is a normalizing constant, and $\beta$ and $\gamma$ are two adjustable parameters.  The  peak frequency occurs at $\omega_{\beta,\gamma}=(\beta/\gamma)^{1/\gamma}$, and we choose $a_{\beta,\gamma}\equiv  2 (e\gamma/\beta)^{\beta/\gamma}$ in order to meet the convention $\Psi_{\beta,\gamma}(\omega_{\beta,\gamma})=2$.   For the generalized Morse wavelets, the duration takes the simple form $P_{\beta,\gamma}=\sqrt{\beta\gamma}$. In \cite{lilly09-itsp}, the $\gamma=3$ family is recommended as a superior alternative to the only approximately analytic Morlet wavelet.  With this choice, the wavelet duration $P_{\beta,\gamma}$ is matched to the signal variability by adjusting $\beta$. It is also shown in  \cite{lilly09-itsp} that time decay of the generalized Morse wavelets is controlled by $\beta$, with $|\psi_{\beta,\gamma}(t)|/|\psi_{\beta,\gamma}(0)|\sim |t|^{-(\beta+1)}$.  Thus to ensure square integrability in (\ref{wavelettransformexpansion}), the constraint $r_\psi> 3$ translates to $\beta>2$.  

\subsection{Ridge Application}
  The multivariate wavelet ridge analysis method using the generalized Morse wavelets is applied to the data set shown in Fig.~\ref{bandwidth_loopers}a, using a freely distributed software package described in Appendix~\ref{appendixjlab}.    For all but two of the time series, the $\gamma=3$, $\beta=3$ generalized Morse wavelets are used, so in this case we have $P_{\beta,\gamma}= \sqrt{\beta\gamma}=3$.  The wavelet transform vector $\bw_{\bx,\psi}(t,s)$ is computed with 82 logarithmically spaced frequency levels, with a lowest frequency of 0.01 cpd (cycles per day) and a maximum frequency of 0.28 cpd.  For two of the time series, the frequency content was at considerably higher frequencies, and so we use different settings in order to more closely approach the Nyquist frequency.  For these two time series, we used the $\gamma=3$, $\beta=8$ generalized Morse wavelets, so $P_{\beta,\gamma}=2\sqrt{6}$, and computed the wavelet transform at 140 logarithmically spaced levels with a lowest frequency of 0.01 cpd and a maximum frequency of 0.34 cpd.

The multivariate ridge method described in Section~\ref{estimationsection} is then applied, rejecting ridges with that execute a smaller number of complete cycles than $2P_{\beta,\gamma}=6$.  At a very small number of points, two valid ridges are obtained which overlap, and these are combined into a single estimated signal $\widehat\bx_\psi(t)$ through a power-weighted average.  Thus there is either one or zero estimated modulated oscillations $\widehat\bx_\psi(t)$ present at each time.   These modulated bivariate oscillations can be converted into the parameters of a time-varying ellipse following \cite{lilly10-itsp}, and these ellipses are shown in Fig.~\ref{bandwidth_loopers}b.    The time interval between successive ellipses is proportional to the estimated period $2\pi/\widehat\omega_{\bx}(t)$, and the ellipses are shown at twice actual size.   A set of estimated residuals formed by subtraction, $\widehat\bx_r(t)\equiv \bx_o(t)-\Re\{\widehat\bx_\psi(t)\}$, shown in Fig.~\ref{bandwidth_loopers}c.  Finally, the estimate bias is shown in Fig.~\ref{bandwidth_loopers}d by converting the estimated deviation $\frac{1}{2}P_\psi^2\widehat{\widetilde\bx}_{2;\psi}(t)$ into time-varying ellipse parameters.  That the estimated bias is generally small compared to the estimated signals is consistent with visual inspection of Fig.~\ref{bandwidth_loopers}c, in which the residual curves appear to be largely devoid of oscillatory motions.

\section{Conclusion}\label{conclusionsection}
This paper has addressed the analysis of modulated oscillations in multivariate time series.  The key contribution is a local expansion of modulated oscillatory variability in terms of deviations from a pure oscillation at a common but time-varying frequency.  This model captures the essence of time-dependent wavelike motion spanning multiple signal channels.  A condition for a signal to be considered a {\em modulated multivariate oscillation} is given, which amounts to demanding that the magnitude of the local deviation of the signal from a pure oscillation is no larger than the magnitude of the signal itself.  

A generalization of wavelet ridge analysis for multivariate timeseries is presented which enables an estimate of the modulated oscillation to be formed from a wavelet transform of the signal.  By appealing to the signal model, constraints may be placed on the choice of analyzing wavelet such that the estimate of a modulated oscillation is guaranteed to have small bias.  By considering signals which are both multivariate as well as non-negligibly modulated, and by presenting forms for an important source of time-varying error, this work substantially extends  earlier tools for analysis of  nonstationary or modulated oscillations.

\appendices

\section{A Freely Distributed Software Package}\label{appendixjlab}
All software associated with this paper is distributed as a part of a freely available Matlab toolbox called Jlab, written by the first author and available at \url{http://www.jmlilly.net}.  The Jsignal module of Jlab includes numerous routines for multivariate wavelet ridge analysis suitable for large data sets. The wavelet transform using generalized Morse wavelets is implemented with \texttt{wavetrans}, which  calls \texttt{morsewave} to compute the wavelets.  The standard univariate and joint wavelet ridges are found by \texttt{ridgewalk} using a numerically efficient algorithm that includes quadratic interpolation between discrete scale levels. Position records given in latitude and longitude are converted into displacement velocities with \texttt{latlon2uv}, while and \texttt{latlon2xy} and \texttt{xy2latlon} convert between latitude and longitude and a local Cartesian coordinate system.  Ellipse parameters are found from a pair analytic of signals with  \texttt{ellparams}, and the ellipses can be plotted using \texttt{ellipseplot}.  Finally, \texttt{makefigs\!\_\,multivariate} generates all figures in this paper.

\section{Expansion of the Rate of Change of the Signal}\label{appendixsignalderivative}
In this appendix a version of the local modulation expansion  (\ref{simplexomegax}) for the first time derivative of the signal is derived.   The partial derivative with respect to the global time $t$ of $\bx_+(t+\tau)$, assumed thrice differentiable, may be expanded as
\begin{multline}\label{simplexderiv}
\frac{\partial }{\partial t}\bx_+(t+\tau)=e^{i\omega_\bx(t)\tau}
\left\{\bx_+'(t)+\tau\left[\widetilde\bx_2(t)+i\omega_\bx(t)\widetilde\bx_1(t)\right] \right.\\\left.+\frac{1}{2}\tau^2\left[\widetilde\bx_3(t)+i\omega_\bx(t)\widetilde\bx_2(t)\right]   +i\frac{1}{2}\omega_\bx'(t)\tau^3\widetilde\bx_2(t)\right.\\\left.+\bm{\epsilon}_{\bx;t}(t,\tau)\right\}
\end{multline}
where $\bm{\epsilon}_{\bx;t}(t,\tau)$ is a remainder term.  To derive this, write the partial $t$-derivative of $\bx_+(t+\tau)$ as \begin{multline}\label{simplexderivfirst}
\frac{\partial }{\partial t}\bx_+(t+\tau)=
\frac{\partial }{\partial t} \left\{  e^{i\omega_\bx(t)\tau}  \left[e^{-i\omega_\bx(t)\tau} \bx_+(t+\tau)\right]\right\}=\\
e^{i\omega_\bx(t)\tau}\frac{\partial }{\partial t}  \left[e^{-i\omega_\bx(t)\tau} \bx_+(t+\tau)\right] +i\tau \omega_\bx'(t)\,\bx_+(t+\tau).
\end{multline}
Substituting from (\ref{simplexomegax}), the term in square brackets on the second line can be Taylor-expanded in $\tau$ as
\begin{multline}\label{simplexderivfirst2}
\frac{\partial }{\partial t}  \left[e^{-i\omega_\bx(t)\tau} \bx_+(t+\tau)\right] \\
= \bx_+'(t) + \tau \widetilde\bx_1'(t) +\frac{1}{2}\tau^2  \widetilde\bx_2'(t)+
\bm{\epsilon}_{\bx'}(t,\tau)
\end{multline}
where the residual takes the form  \cite[p~880]{abramowitz}
  \begin{equation}
\bm{\epsilon}_{\bx'}(t,\tau)\equiv \frac{1}{6}\tau^3\frac{\partial^3}{\partial\tau^3}\left.\left\{ \frac{\partial}{\partial t} \left[e^{-i\omega_\bx(t)\tau}\bx_+(t+\tau)\right]\right\}\right|_{\tau=v}
 \end{equation}
for some (unknown) point $v$ contained in the interval $\left[0,\tau\right]$.  Note that $v$ is not in general the same as the point $u$ appearing in the expression (\ref{brdef}) for the remainder $\bm{\epsilon}_\bx(t,\tau)$ in the comparable expansion of $\bx_+(t)$.  The identities
\begin{align}
\widetilde\bx_1'(t) +i\omega_\bx'(t)\bx_+(t)& = \widetilde\bx_2(t) + i \omega_\bx(t) \widetilde\bx_1(t) \\
\widetilde\bx_2'(t) +2i\omega_\bx'(t)\widetilde\bx_1(t) & = \widetilde\bx_3(t) + i \omega_\bx(t) \widetilde\bx_2(t)  \label{xtilde2deriv}
\end{align}
 may readily be verified from the definitions (\ref{curvdef}) of the deviation vectors. Substituting these into (\ref{simplexderivfirst2}), and combining the result into (\ref{simplexderivfirst}) together with the local  expansion  in $\tau$  of $\bx_+(t+\tau)$ given by (\ref{simplexomegax}),  we obtain  (\ref{simplexderiv}) with
 \begin{equation}
\bm{\epsilon}_{\bx;t}(t,\tau)\equiv\bm{\epsilon}_{\bx'}(t,\tau)+i\tau\omega_\bx'(t)\bm{\epsilon}_\bx(t,\tau)
 \end{equation}
 as the form of the remainder term.  In the above, we have taken care to avoid differentiating a remainder term such as $\bm{\epsilon}_\bx(t,\tau)$.

 \section{Time Derivative of the Wavelet Transform}\label{appendixfrequency}
Here an expression for the time derivatives of the wavelet transform is found that is valid near the instantaneous frequency curve, using the expansion (\ref{simplexderiv}) for the rate of change of the signal derived in the previous appendix. The normalized time derivative of the wavelet transform is found to be
\begin{multline}\label{derivexpress}
\frac{\frac{\partial }{\partial t}\bw_{\bx,\psi}(t,s) }{\omega_\bx(t)} =\frac{1}{\omega_\bx(t)}\frac{1}{2} \int_{-\infty}^{\infty}
  \!\psi^*(\tau)\,\frac{\partial}{\partial t}\bx_+(t+s\tau)\,d  \tau\\
 =
\Phi_0(t,s)\frac{\bx_+'(t)}{\omega_\bx(t)}
-i\Phi_1(t,s)\left[\frac{\widetilde\bx_2(t)}{\omega_\bx^2(t)}+i\frac{\widetilde\bx_1(t)}{\omega_\bx(t)}\right]
\\-\frac{1}{2}\Phi_2(t,s)\left[\frac{\widetilde\bx_3(t)}{\omega_\bx^3(t)}+i\frac{\widetilde\bx_2(t)}{\omega_\bx^2(t)}\right]
-\frac{1}{2}\Phi_3(t,s)\frac{\omega_\bx'(t)}{\omega_\bx^2(t)}\frac{\widetilde\bx_2(t)}{\omega_\bx^2(t)}
 \\+\Delta\bw_{\bx,\psi;t}(t,s)\end{multline}
by inserting (\ref{simplexderiv}) into the time derivative of  (\ref{simplew}), exchanging the orders of differentiation and integration, and making use of the  definition  (\ref{fdef})  of the $\Phi_p(t,s)$ functions.   The residual term $\Delta\bw_{\bx,\psi;t}(t,s)$ here is again implicitly defined as the difference between the left-hand side and the other terms on the right-hand side; see Appendix~D of \cite{lilly10-itit} for details on bounding this term.  Gathering orders in (\ref{derivexpress}), we find
\begin{multline}\label{transinstfreq}
\frac{\frac{\partial }{\partial t}\bw_{\bx,\psi}(t,s) }{\omega_\bx(t)}
  =\overset{O(1)}{\overbrace{ \frac{\bx_+'(t)}{\omega_\bx(t)}}}
- \overset{O(\delta_T)}{\overbrace{ i\frac{1}{2}\widetilde\Psi_2^*(\omega_\psi)\frac{\widetilde\bx_2(t)}{\omega_\bx^2(t)}}}\\
+ \overset{O(\delta_T^2)}{\overbrace{\Delta\omega_\psi(t,s)\widetilde\Psi_2^*(\omega_\psi) \frac{\widetilde\bx_1(t)}{\omega_\bx(t)}
 -\frac{1}{2}\widetilde\Psi_2^*(\omega_\psi)\frac{\widetilde\bx_3(t)}{\omega_\bx^3(t)}}}  \\
+O(\delta_T^3)
+\Delta\bw_{\bx,\psi;t}(t,s)
\end{multline}
making use (\ref{f})--(\ref{f4}) together with fact that $\omega_\bx'(t)/\omega_\bx^2(t)$ is $O(\delta_T^2)$.  This implies that the first derivative of the signal $\bx_+'(t)$ is accurately estimated by the value of the partial time derivative of  $\bw_{\bx,\psi}(t,s)$  evaluated along the ridge curve.

For what follows, assume that the wavelets are real-valued in the frequency domain. Note that in the instantaneous frequency neighborhood the imaginary part of the projection of the wavelet transform onto its own time derivative is
\begin{multline}\label{wwprime}
\frac{1}{\omega_\bx(t)} \frac{\Im\left\{\bw_{\bx,\psi}^H(t,s)\frac{\partial}{\partial t}\bw_{\bx,\psi}(t,s)\right\}}{\|\bx_+(t)\|^2}=
\\ 1-\overset{O(\delta_T)}{\overbrace{\widetilde\Psi_2(\omega_\psi)\frac{1}{\omega_\bx^2(t)} \frac{\Re\left\{\bx_+^H(t)\widetilde\bx_2(t)\right\} }{\|\bx_+(t)\|^2}}} +\overset{O(\delta_T^2)}{\overbrace{\frac{1}{4}\widetilde\Psi_2^2(\omega_\psi) \frac{\xi_\bx^2(t)}{\omega_\bx^4(t)}}}\\
 +\overset{O(\delta_T^2)}{\overbrace{\frac{1}{2}\widetilde\Psi_2(\omega_\psi)\frac{1}{\omega_\bx^3(t)}
\left[ \frac{\Im\left\{\widetilde\bx_1^H(t)\widetilde\bx_2(t)-\bx_+^H(t)\widetilde\bx_3(t)\right\}}{\|\bx_+(t)\|^2}\right] }}
  \\
 + O(\delta_T^3) +  O\left(\frac{\Delta\bw_{\bx,\psi}(t,s)}{\|\bx_+(t)\|^2}\right) +   O\left(\frac{\Delta\bw_{\bx,\psi;s}(t,s)}{\|\bx_+(t)\|^2} \right)
\end{multline}
as  follows by combining the two expansions (\ref{readytoexpand3}) and (\ref{transinstfreq}), and using the substitution $\bx_+'(t)=\widetilde\bx_+(t)+i\omega_\bx(t)\bx_+(t)$; note that the order $\delta_T$ term in (\ref{wwprime}) arises twice, and thus its coefficient is unity rather than $1/2$.   At the same time we may find, again for real-valued wavelets,
\begin{multline}\label{ww}
 \frac{\|\bw_{\bx,\psi}(t,s)\|^2}{\|\bx_+(t)\|^2}=
1 - \overset{O(\delta_T)}{\overbrace{\widetilde\Psi_2(\omega_\psi)\frac{1}{\omega_\bx^2(t)} \frac{\Re\left\{\bx_+^H(t)\widetilde\bx_2(t)\right\} }{\|\bx_+(t)\|^2}}}
 \\+\overset{O(\delta_T^2)}{\overbrace{\frac{1}{4}\widetilde\Psi_2^2(\omega_\psi)\frac{\xi_\bx^2(t)}{\omega_\bx^4(t)}}}
 + O(\delta_T^3) +O\left(\frac{\Delta\bw_{\bx,\psi}(t,s)}{\|\bx_+(t)\|^2}\right)
 \end{multline}
for the expansion of the modulus-squared wavelet transform in the instantaneous frequency neighborhood.   In deriving both of these expressions we have made use of the fact that $\bx_+^H(t)\widetilde\bx_1(t)$ is purely real, see (\ref{parallelfirstderivreal}).

The transform instantaneous frequency in the $\delta_T^2$~neighborhood of the signal's instantaneous frequency is then given by
\begin{multline}\label{freqexpand}
\frac{\Omega_\psi(t,s)}{\omega_\bx(t)} \equiv \frac{1}{\omega_\bx(t)}\frac{\Im\left\{\bw_{\bx,\psi}^H(t,s)\frac{\partial}{\partial t}\bw_{\bx,\psi}(t,s)\right\}}{\|\bw_{\bx,\psi}(t,s)\|^2}=\\
1 +\overset{O(\delta_T^2)}{\overbrace{\frac{1}{2}\widetilde\Psi_2(\omega_\psi)\frac{1}{\omega_\bx^3(t)}
\left[ \frac{\Im\left\{\widetilde\bx_1^H(t)\widetilde\bx_2(t)-\bx_+^H(t)\widetilde\bx_3(t)\right\}}{\|\bx_+(t)\|^2}\right] }}\\
 + O(\delta_T^3) +O\left(\frac{\Delta\bw_{\bx,\psi}(t,s)}{\|\bx_+(t)\|^2}\right) +   O\left(\frac{\Delta\bw_{\bx,\psi;s}(t,s)}{\|\bx_+(t)\|^2} \right)
 \end{multline}
which is found by combining (\ref{wwprime}) and (\ref{ww}), using $1/(1+x)=1-x+x^2+O(x^3)$, and carefully keeping track of the cross terms from the product of the two expansions.  A number of cancelations occur:  the leading-order term in the numerator cancels the leading-order term in the denominator, and the square of the first term in (\ref{ww}) (arising from the expansion of the denominator) cancels an identical term arising from the product of the numerator with the expansion of the denominator. The result (\ref{freqexpand}) implies that the transform frequency evaluated along the ridge, $\widehat\omega_\psi(t)\equiv \Omega_\psi(t,\widehat s(t))$, is an accurate estimate of the joint instantaneous frequency $\omega_\bx(t)$.

\section{Scale Derivative of the Wavelet Transform}\label{appendixscalederiv}

The scale derivative of the wavelet transform can be found in a similar fashion to the time derivative in the preceding appendix.   The scale derivative of the shifted analytic signal $\bx_+(t+s\tau)$ is related to its time derivative via
\begin{equation}
s\frac{\partial }{\partial s}\bx_+(t+s\tau)=
s\tau\frac{\partial }{\partial t}\bx_+(t+s\tau).
\end{equation}
Then inserting (\ref{simplexderiv}) into the scale derivative of  the wavelet transform expression (\ref{simplew}) and again using  (\ref{fdef}) leads to
\begin{multline}\label{scaleexpress}
s\frac{\partial }{\partial s}\bw_{\bx,\psi}(t,s)  =\frac{1}{2}\int_{-\infty}^{\infty}
  \!s\tau\psi^*(\tau)\,\frac{\partial}{\partial t}\bx_+(t+s\tau)\,d  \tau\\
 =
-i\Phi_1(t,s)\frac{\bx_+'(t)}{\omega_\bx(t)}
-\Phi_2(t,s)\left[\frac{\widetilde\bx_2(t)}{\omega_\bx^2(t)}+i\frac{\widetilde\bx_1(t)}{\omega_\bx(t)}\right]
\\+i\frac{1}{2}\Phi_3(t,s)\left[\frac{\widetilde\bx_3(t)}{\omega_\bx^3(t)}+i\frac{\widetilde\bx_2(t)}{\omega_\bx^2(t)}\right]
+i\frac{1}{2}\Phi_4(t,s)\frac{\omega_\bx'(t)}{\omega_\bx^2(t)}\frac{\widetilde\bx_2(t)}{\omega_\bx^2(t)}
 \\+\Delta\bw_{\bx,\psi;s}(t,s)\end{multline}
with the residual $\Delta\bw_{\bx,\psi;s}(t,s)$ defined implicitly as before.  Again, we refer the reader to Appendix~D of \cite{lilly10-itit} for details on bounding this term.   Noting  (\ref{f})--(\ref{f5}) and using $\bx_+'(t)=\widetilde\bx_1(t)+i\omega_\bx(t)\bx_+(t)$, we can gather orders in (\ref{scaleexpress}) to yield
\begin{multline}
s\frac{\partial }{\partial s}\bw_{\bx,\psi}(t,s) \label{scaleresult}
  = \overset{O(1)}{\overbrace{-i\Psi_2^*(\omega_\psi)\frac{\widetilde\bx_1(t)}{\omega_\bx(t)}}}\\
\overset{O(\delta_T)}{\overbrace{+\Delta\omega_{\bx,\psi}(t,s)\widetilde\Psi_2^*(\omega_\psi)\bx_+(t)-\left[\widetilde\Psi_2^*(\omega_\psi)+\frac{1}{2}\widetilde\Psi_3^*(\omega_\psi)\right]\frac{\widetilde\bx_2(t)}{\omega_\bx^2(t)}}}\\+O(\delta_T^2) +\Delta\bw_{\bx,\psi;s}(t,s)
\end{multline}
in which terms up to first order in $\delta_T$ are resolved.  This implies that a suitably normalized version of the scale derivative of the wavelet transform evaluated along the ridge recovers the first intrinsic deviation vector $\widetilde\bx_1(t)$.

The ridge condition can now be evaluated using expressions for the wavelet transform and its scale derivative.  The ridge condition $\frac{\partial}{\partial s}\|\bw_{\bx,\psi}(t,s)\|=0$, from (\ref{ampvectorridge}), is equivalent to
\begin{equation}
\Re\left\{\frac{\bw_{\bx,\psi}^H(t,s) \left[s\frac{\partial }{\partial s}\bw_{\bx,\psi}(t,s) \right]}{\|\bx_+(t)\|^2}\right\}=0
\end{equation}
and  inserting (\ref{readytoexpand3}) and (\ref{scaleresult}), we find
\begin{multline}
\frac{\bw_{\bx,\psi}^H(t,s) \left[s\frac{\partial }{\partial s}\bw_{\bx,\psi}(t,s) \right]}{\|\bx_+(t)\|^2}=\label{bigexpression}
\overset{O(1)}{\overbrace{-i\widetilde\Psi_2^*(\omega_\psi)\frac{1}{\omega_\bx(t)}\frac{\bx_+^H(t)\widetilde\bx_1(t)}{\|\bx_+(t)\|^2}}}
\\\overset{O(\delta_T)}{\overbrace{+\Delta\omega_{\bx,\psi}(t,s)\widetilde\Psi_2^*(\omega_\psi)+i\frac{1}{2}\left[\widetilde\Psi_2^*(\omega_\psi)\right]^2
\frac{1}{\omega_\bx^3(t)}\frac{\widetilde\bx_2^H(t)\widetilde\bx_1(t)}{\|\bx_+(t)\|^2}}}\\\overset{O(\delta_T)}{\overbrace{-\left[\widetilde\Psi_2^*(\omega_\psi)+\frac{1}{2}\widetilde\Psi_3^*(\omega_\psi)\right]\frac{1}{\omega_\bx^2(t)}\frac{\bx_+^H(t)\widetilde\bx_2(t)}{\|\bx_+(t)\|^2}}}\\
+O(\delta_T^2) +  O\left(\frac{\Delta\bw_{\bx,\psi}(t,s)}{\|\bx_+(t)\|^2}\right) +   O\left(\frac{\Delta\bw_{\bx,\psi;s}(t,s)}{\|\bx_+(t)\|^2} \right).
\end{multline}
Assuming that the wavelets are real-valued, setting the real part of (\ref{bigexpression}) equal to zero leads to
\begin{multline}
\Delta\omega_{\bx,\psi}(t)=\left[1+\frac{1}{2}\frac{\Psi_3(\omega_\psi)}{\Psi_2(\omega_\psi) }\right] \frac{1}{\omega_\bx^2(t)}\frac{\Re\left\{\bx_+^H(t)\widetilde\bx_2(t)\right\}}{\|\bx_+(t)\|^2}\\
+\frac{1}{2}\Psi_2(\omega_\psi)\frac{1}{\omega_\bx^3(t)}\frac{\Im\left\{\widetilde\bx_2^H(t)\widetilde\bx_1(t)\right\}}{\|\bx_+(t)\|^2}\\+O(\delta_T^3)+O\left(\frac{\Delta\bw_{\bx,\psi}(t,s)}{\|\bx_+(t)\|^2}\right) +   O\left(\frac{\Delta\bw_{\bx,\psi;s}(t,s)}{\|\bx_+(t)\|^2} \right)
\end{multline}
along a ridge.  Note that the leading order term in $ \Delta\omega_{\bx,\psi}(t)$ is second order in $\delta_T$ along the ridge, in agreement with the assumption that $s$ lies within the  instantaneous frequency neighborhood.  In terms of scale, the ridge curve is then given from (\ref{neighborhooddef}) by $\widehat s(t) = \left[1+\Delta\omega_{\bx,\psi}(t)\right]\omega_\psi/\omega_\bx(t)$.

\begin{biography}{Jonathan M. Lilly}
(M05) was born in Lansing, Michigan, in 1972. He received the B.S. degree in geology and geophysics from Yale University, New Haven, Connecticut, in 1994, and the M.S. and Ph.D. degrees in physical oceanography from the University of Washington (UW), Seattle, Washington, in 1997 and 2002, respectively. 

He was a Postdoctoral Researcher with the UW Applied Physics Laboratory and School of Oceanography, from 2002 to 2003, and with the Laboratoire d'Oc\'eanographie Dynamique et de Climatologie, Universit\'e Pierre
et Marie Curie, Paris, France, from 2003 to 2005.  From 2005 until 2010,  he was a Research Associate with Earth and Space Research in Seattle, Washington.  In 2010 he joined NorthWest Research Associates, an employee-owned scientific research corporation in Redmond, Washington,  as a Senior Research Scientist.  His research interests are oceanic vortex structures,  time/frequency analysis methods, satellite oceanography, and wave--wave interactions. 

Dr. Lilly is a member of the American Meteorological Society and of the American Geophysical Union.
\end{biography}

\begin{biography}{Sofia C. Olhede} was born in Spanga, Sweden, in 1977.
She received the M. Sci. and Ph.D. degrees in mathematics from Imperial
College London, London, U.K., in 2000 and 2003, respectively.

She held the posts of Lecturer (2002--2006) and Senior Lecturer
(2006--2007) with the Mathematics Department, Imperial College London, and
in 2007, she joined the Department of Statistical Science, University
College London, where she is Professor of Statistics and director of
research.  She holds a UK Engineering and Physical Sciences Research
Council Leadership fellowship in Statistics. Her research interests
include the analysis of complex-valued stochastic processes,
non-stationary time series and inhomogeneous random fields, with
applications in neuroscience and oceanography.

Prof. Olhede is an Associate Editor of IEEE Transactions on Signal
Processing.
\end{biography}

\label{lastpage}

\end{document}